\documentclass[useAMS,usenatbib]{mn2e}
\usepackage{epsfig,float,natbib,amsmath,longtable,supertabular}

\title[Monthly Notices: \LaTeXe\ guide for authors]
  {Monthly Notices of the Royal Astronomical
  Society: \\ \LaTeXe\ style guide for authors}

  \title[First Results from the CHEPS]
  {First Results from the CHEPS\thanks{Calan-Hertfordshire Extrasolar Planet Search}: Exoplanets and the Discovery of an Eccentric Brown Dwarf in the Desert\thanks{Based on observations collected at the La Silla Paranal Observatory, ESO (Chile) with the HARPS spectrograph on the ESO 3.6m telescope, under the program IDs 
079.C-0927(B), 079.C-0927(C), 081.C-0148(A), 081.C-0148(B) and 282.C-5034.}}
\author[J. S. Jenkins et al.]
  {J.S. Jenkins$^{1}$, H.R.A. Jones$^{2}$, K. Go{\'z}dziewski$^{3}$, C. Migaszewski$^{3}$, J.R. Barnes$^{2}$, \\ 
  \newauthor M.I. Jones$^{1}$, P. Rojo$^{1}$, D.J. Pinfield$^{2}$, A.C. Day-Jones$^{2}$, S. Hoyer$^{1}$\\
  $^1$Department of Astronomy, Universidad de Chile, Casilla 36-D, Santiago, Chile, email: jjenkins@das.uchile.cl \\
  $^2$Center for Astrophysics, University of Hertfordshire, College Lane Campus, Hatfield, Hertfordshire, UK, AL10 9AB\\
  $^3$Toru{\'n} Centre for Astronomy, Nicolaus Copernicus University, Gagarina 11, 87-100 Toru{\'n}, Poland}

\date{Draft: 11/08}

\pagerange{\pageref{firstpage}--\pageref{lastpage}} \pubyear{2008}

\def\LaTeX{L\kern-.36em\raise.3ex\hbox{a}\kern-.15em
    T\kern-.1667em\lower.7ex\hbox{E}\kern-.125emX}

\begin{document}

\label{firstpage}

\maketitle

\begin{abstract}

We report the discovery of a brown dwarf on an eccentric orbit and with a semimajor axis that places it in the brown dwarf desert region around the star HD191760.  The star has a spectral type of 
G3IV/V and a metallicity ([Fe/H]) of 0.29~dex.  HD191760 adds to the small number of metal-rich stars with brown dwarf 
companions.  The brown dwarf (HD191760$b$) is found to have an orbital period of 505.57$\pm$0.40~days and semimajor axis of 1.35$\pm$0.01~AU, placing it 
firmly in the brown dwarf desert.  The eccentricity of HD191760$b$ is found to be 0.63$\pm$0.01, meaning it reaches as close as 0.5~AU from 
the host star.  Dynamical simulations indicate that no inner planets could reside at separations 
beyond $\sim$0.17~AU due to the disastrous gravity imposed by HD191760$b$.  In addition to these first results we also refine the orbits found for the exoplanets around the stars HD48265, 
HD143361 and HD154672.  All 1-planet solutions are in agreement with those previously published by the Magellan Planet Search.  

\end{abstract}
\begin{keywords}

stars:~low-mass, brown dwarfs -- planetary systems -- techniques: radial velocities 

\end{keywords}

\section{Introduction}

Since the first few extrasolar planets (hereafter exoplanets) were detected (\citealp{mayor95}; \citealp{marcy96a}; \citealp{marcy96b}) it became readily apparent that planets preferred 
metal-rich environments (\citealp{gonzalez}).  This discovery has been placed in the framework of giant planet formation through core accretion of gas depleted materials left over from 
the formation of the parent star (e.g. \citealp{kornet05}; \citealp{alibert05}).  This hypothesis has been extensively tested by different 
methods and analysis techniques, most recently by \citet{fischer05}, and 
remains the most plausible scenario for the increased probability of planet detection around metal-rich stars.  The competing theory is based on the 
concept that planets migrate through the proto-planetary disks from which they formed (\citealp{lin}; \citealp{trilling02}).  It has been hypothesized that the stellar atmosphere is polluted after 
formation through infall of migrating planets onto the stellar surface.  However, \citet{pinsonneault01} modelled the infall of material onto stellar atmospheres at the ZAMS and found 
that 10~M$_{\rm{\oplus}}$ worth of gas depleted material could raise the metallicities ([Fe/H]) of G dwarfs to $\sim$0.3~dex, whereas the same material falling on an F dwarf would 
increase the metallicity to $\sim$0.6~dex.  In their large sample ($>$1000 stars) of consistent metallicities, Fischer \& Valenti found no correlation between the depth of convective zones 
of exoplanet hosts and their metallicity, indicating that infall of planets or gas depleted material can not explain the over abundance of metal-rich atmospheres for 
exoplanet host stars.  

Whatever the scenario that leads to exoplanet host stars being rich in metals, it is clear that such stars have a higher probability of hosting 
exoplanets.  Fischer \& Valenti have determined that the fraction of planets increases with increasing metallicity proportional to the square of the 
number of iron atoms in the stellar photosphere.  This correlation is being 
exploited by a number of planet search projects like the N2k, Keck, etc who bias their samples, or subsets thereof, towards the most metal-rich stars.  
We are exploiting this 
correlation also, by searching around the most metal-rich solar-type stars in the southern hemisphere as part of our CHEPS project.  The CHEPS target 
list is compiled from the analysis of \citet{jenkins08} 
by utilising high S/N high resolution spectra, acquired using the FEROS spectrograph (\citealp{kaufer99}).  The targets were selected to have [Fe/H]$\ge$0.1~dex and all 
have $B-V$ colours in the range 0.5$-$0.9 (Spectral Types of late-F to mid-K).  Along with the metallicity bias we are also focusing on the most inactive stars.  The Ca \sc ii \rm HK 
line cores were used in Jenkins et al. to measure the log$R'_{\rm{HK}}$ activity index (\citealp{noyes}) which has been shown to be a useful proxy to the level of 
noise (aka $jitter$) in radial-velocity timeseries (\citealp{saar,saar98}; \citealp{wright05}).  We therefore selected our primary sample to include all stars with 
log$R'_{\rm{HK}}$$\le$-4.5, which allows for the lowest levels of radial-velocity jitter in our dataset.

\section{Observations}

We present data from six observing runs using the ESO-HARPS instrument (\citealp{mayor03b}) on the 3.6m telescope in La Silla Chile.  HARPS is a 
fiber-fed 
cross-dispersed echelle spectrograph.  The instrument employs two 1$''$ on sky fibers, one on object and one on the Thorium-Argon (ThAr) calibration 
source, which feed the light to the spectrograph to be re-imaged on two 2k4 CCD chips.  A total of 72 orders are spread over the two CCD's, 36 object and 36 
calibration source, with an instrumental resolution ($R$) of $\sim$115,000.  29 observations were 
made of HD191760, with 8 for HD48265 and 6 each for both HD143361 and HD154672, all using our standard observational procedure, which aims to observe 
each star for a fixed integration time of 15~minutes in order to integrate over 
the strongest p-mode oscillations that all solar-like stars exhibit (\citealp{o'toole08}).  Also to attempt to increase our efficiency to the detection 
of short period exoplanets, and in particular low mass exoplanets, we intensively monitor the stars on short timescales to search for potential planetary signals, including for HD191760.

The reduction and analysis of all stars are performed in near real time at the telescope, using the latest version of the HARPS-DRS (\citealp{pepe04}).  
The DRS (Data Reduction Software) performs all functions 
of the reduction and analysis, from bias removal, order localisation, flatfielding, cosmic ray removal, scattered light subtraction, extraction and 
blaze correction, to determination of the radial-velocities, barycentric Julian dates and internal uncertainties through cross-correlation with stellar 
templates and the 
reference ThAr calibration orders.  Due to the high quality mechanical, thermal and pressure stabilisation of HARPS, 
the 
drift over a single night is found to be less than 1~m/s.  Such a system provides internal uncertainties below 1~m/s, meaning our short term precision is 
dominated by stellar activity induced jitter (e.g. \citealp{saar98}).

\section{HD191760 Characteristics}

Table~\ref{tab:params} lists the stellar characteristics of the star HD191760.  \citet{houk78} lists the star's spectral type as G3IV/V, which is in agreement with the value 
adopted by Hipparcos (\citealp{perryman}; \citealp{leeuwen08}) and also with the derived distance from the Hipparcos main sequence 
($\Delta M_{V}$; \citealp{jenkins08}).  All Hipparcos astrometry was obtained using the AstroGrid tool which lists the new catalogue values from 
van Leeuwen with increased astrometric precision.  The kinematic catalogue of \citet{nordstrom04}, later updated by \citet{holmberg07}, includes observations of HD191760 through Str{\"o}mgren filters and 
calibrations based on the infrared flux method (\citealp{alonso96}) to derive various stellar characteristics.  They also used parallax measurements from Hipparcos, along with Padova 
evolutionary tracks (\citealp{girardi00}) to determine evolutionary parameters for their sample.  We have also derived various parameters for a number 
of stars in preparation for our metallicity based planet search, both by high resolution FEROS spectra and precise photometry.  We used 
the Hipparcos optical magnitudes, along with 2MASS 
photometry, to measure the effective temperatures of our stars by the infrared flux method (\citealp{blackwell90}).  We used these temperatures to interpolate a grid of high 
resolution Kurucz model atmospheres (\citealp{kurucz93}), built using WITA6 (\citealp{pavlenko00}), to determine the stellar metallicity.  Yonsei-Yale (Y2) isochrones 
(\citealp{demarque04}) were then interpolated using these temperatures, metallicities and an estimate for the alpha-enhancement parameter that was obtained by the locus of points in 
\citet{edvardsson93} and \citet{pagel95}.

\begin{figure}
\vspace{5.0cm}
\hspace{-4.0cm}
\includegraphics{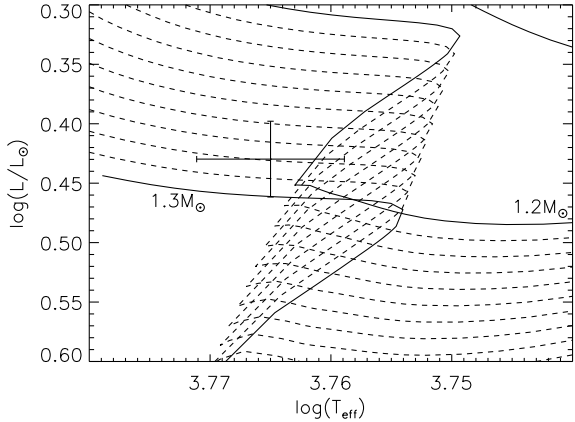}
\vspace{0.5cm}
\caption{Evolutionary Y2 mass tracks are shown in steps of 0.1$M_{\rm{\odot}}$ (solid curves) and finer steps of 0.01$M_{\rm{\odot}}$ (dashed curves) for effective temperatures and 
bolometric luminosities.  HD191760 is represented by the error bars and its position is consistent with a mass of 1.28$^{+0.02}_{-0.10}$M$_{\rm{\odot}}$ and an age of 
4.21$^{+0.5}_{-2.8}$Gyrs.  The masses of the solid curves are shown for reference.}
\label{hr}
\end{figure}

The properties of HD191760 derived by Holmberg et al. and this work mostly agree within the formal uncertainties.  Their metallicity value of 
0.12$\pm$0.10~dex is found to be smaller than our value of 0.29$\pm$0.07~dex and \citet{holmberg08} show that there is a small systematic offset between their metallicities and those published 
in Jenkins et al. due to the differing effective temperature scales employed.  Fig.~\ref{hr} shows the position of HD191760 
on an HR-diagram, along with the Y2 isochrones in steps of 0.1$M_{\odot}$ (solid curves) and higher resolution steps of 0.01$M_{\odot}$ (dashed curves).  The position 
of the star, given the bolometric luminosity of 2.69$\pm$0.2L$_{\odot}$, effective temperature of 5821$\pm$82K and [Fe/H] of 0.29$\pm$0.07~dex, is best 
explained by a star of mass 1.28$^{+0.02}_{-0.10}$$M_{\odot}$, where the uncertainties are dominated by the 
uncertainty on the Hipparcos parallax.  Our estimate for the age of the star is 4.1$^{+0.8}_{-2.8}$Gyrs.  The large lower error bars arise due to the 
position of the star close to the main sequence turnoff.  The surface gravity (log$g$) value of 4.13$^{+0.05}_{-0.04}$dex is consistent with a star 
evolving onto the subgiant branch.

\begin{table}
\center
\caption{Stellar parameters for HD191760.}
\label{tab:params}
\begin{tabular}{cc}
\hline
\multicolumn{1}{c}{Parameter}& \multicolumn{1}{c}{HD191760} \\ \hline

Spectral Type                          & G3IV/V \\
log$R'$$_{\rm{HK}}$                    & -5.17 \\
Hipparcos $N$$_{\rm{obs}}$             & 99 \\
Hipparcos $\sigma$                     & 0.011 \\
$\pi$ (mas)                            & 12.25$\pm$0.90 \\
$\Delta$$M_{V}$                        & 1.107 \\
$L_{\rm{\star}}$/$L_{\odot}$           & 2.69$\pm$0.20 \\
$M_{\rm{\star}}$/$M_{\odot}$ - H07     & 1.14$^{+0.08}_{-0.05}$ \\
$M_{\rm{\star}}$/$M_{\odot}$ - Here    & 1.28$^{+0.02}_{-0.10}$ \\
$R_{\rm{\star}}$/$R_{\odot}$           & 1.62$\pm$0.07 \\
$T$$_{\rm{eff}}$ (K) - H07             & 5794$\pm$76 \\
$T$$_{\rm{eff}}$ (K) - J08             & 5821$\pm$82 \\
$[$Fe/H$]$ - H07                       & 0.12$\pm$0.10 \\
$[$Fe/H$]$ - J08                       & 0.29$\pm$0.07 \\
log($g$)                               & 4.13$^{+0.05}_{-0.04}$ \\
U,V,W - H07 (km/s)                     & -29,-21,15 \\
$v$sin$i$$_{R'_{\rm{HK}}}$ (km/s)      & 3.47 \\
$v$sin$i$$_{\rm{H07}}$ (km/s)          & 2  \\
$v$sin$i$$_{\rm{CCF}}$ (km/s)          & 2.33$\pm$0.05 \\
$P_{\rm{rot}}$$_{,R'_{\rm{HK}}}$ (days)     & 25.2 \\
$P_{\rm{rot}}$$_{,\rm{CCF}}$ (days)     & 35.1 \\
Age$_{R'_{\rm{HK}}}$ (Gyrs)            & 9.9 \\
Age (Gyrs) - H07                       & 5.6$^{+0.9}_{-1.9}$ \\
Age (Gyrs) - Here                      & 4.1$^{+0.8}_{-2.8}$ \\
Jitter - log$R'$$_{\rm{HK}}$ (m/s)     & 1.50 \\

\hline
\end{tabular}

\medskip
H07 relates to \citet{holmberg07}, J08 is the reference for \citet{jenkins08} and Here labels values derived in this work.
\end{table}

\begin{figure}
\vspace{5.0cm}
\hspace{-4.0cm}
\includegraphics{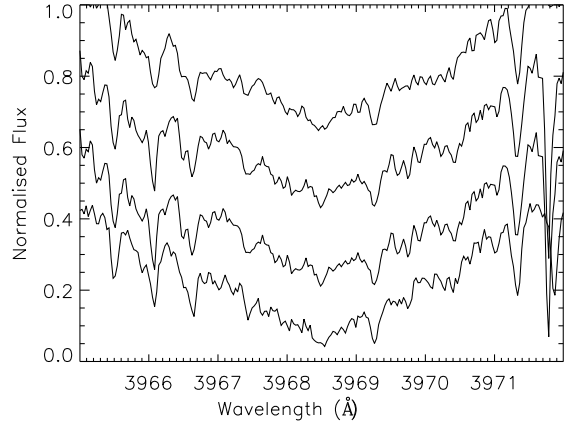}
\vspace{0.5cm}
\caption{The calcium \sc ii \rm H line cores for the four stars discussed in this work.  From top to bottom we show HD191760, HD48265, HD143361 and 
HD154672 respectively.  All four objects exhibit no re-emission of flux in the line core, highlighting the inactive nature of these stars.}
\label{activity}
\end{figure}

As part of the analysis we performed to select our target sample we also extracted chromospheric activity indices from measurements of the level of emission in the 
Ca \sc ii \rm HK lines.  The log$R'$$_{\rm{HK}}$-index has been shown to correlate well with the level of noise (aka jitter) in precise radial-velocity datasets 
(e.g. \citealp{santos00}; \citealp{wright05}).  Fig.~\ref{activity} shows the Ca \sc ii \rm H line core of HD191760 (top) along with the other three 
planet-host stars we present data for in $\S$5.  Our low derived value of log$R'$$_{\rm{HK}}$=-5.17 is highlighted from the deep central line core, or lack of core re-emission, 
of HD191760.  

\section{Orbital Solution for HD191760}

\begin{figure}
\vspace{2cm}
\hspace{-4.0cm}
\includegraphics{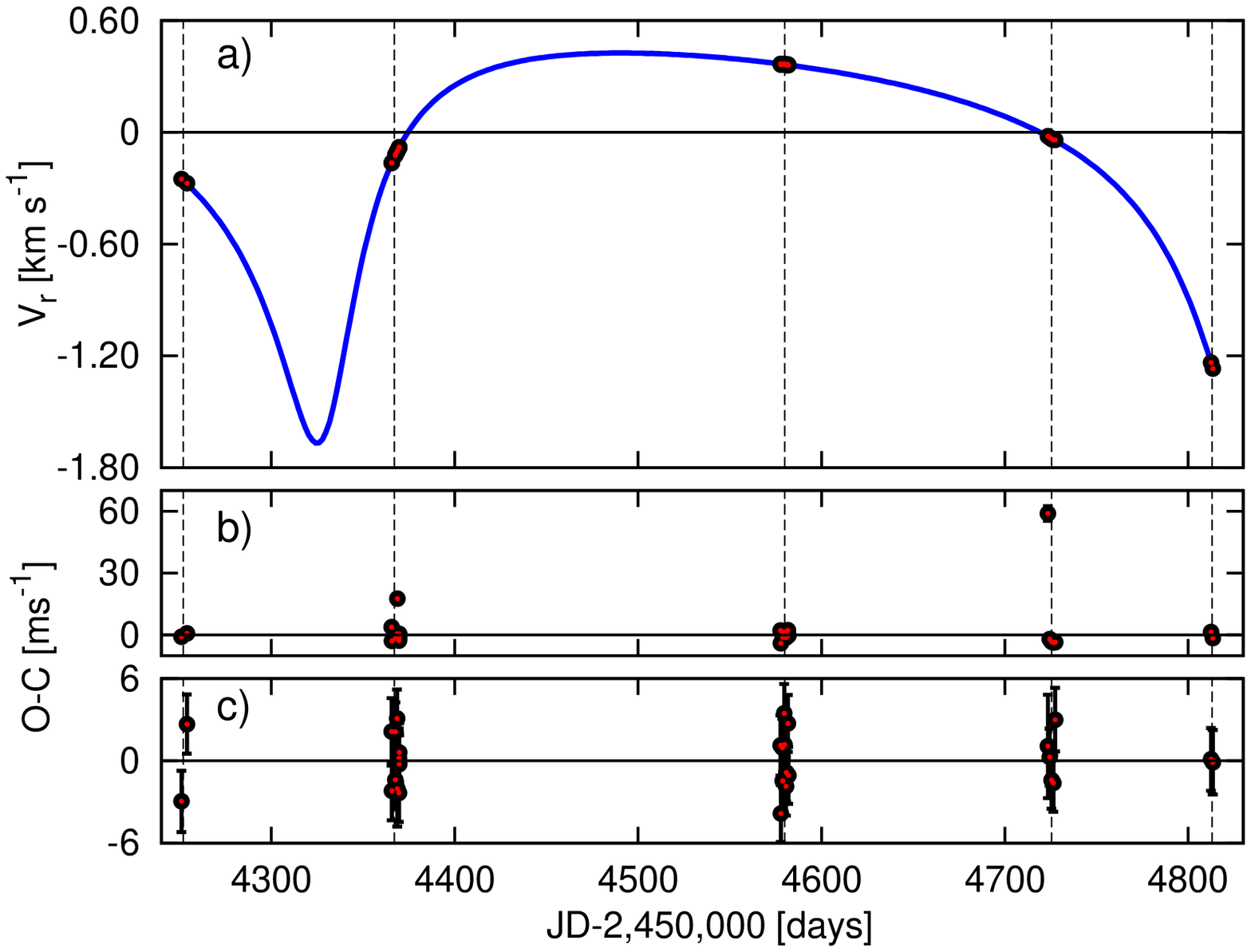}
\includegraphics{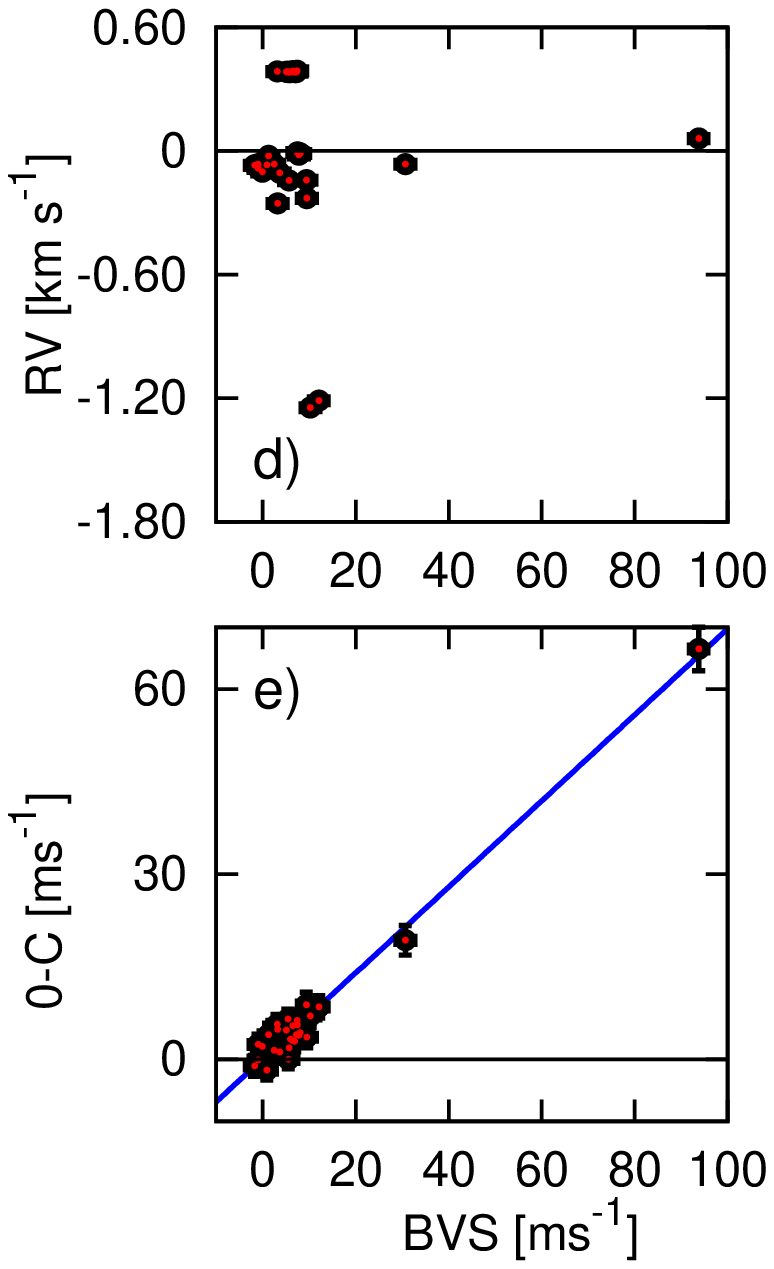}
\vspace{3.5cm}
\caption{Panel (a) shows the best fit Keplerian solution to the Dopper points for HD191760 with BVS correction as in \citet{migaszewski09}.  The data come from six observing runs spread 
over a full orbital period of the companion.  The addition of two Doppler points close to periastron passage was used to confirm the eccentric nature of the companion.  Panel (b) shows 
the residuals without BVS correction and panel (c) shows the residuals after correction, corresponding to the synthetic curve in panel (a).  Panel (d) shows the radial-velocities 
against the BVS after subtraction of the best fit Keplerian synthetic curve.  Panel (e) shows the best fit linear correlation 
coefficient of BVS-RV.  Note that the two values with the largest BVS help to better constrain the correlation.}
\label{fit}
\end{figure}

The 29 Dopper velocities for HD191760 were measured in six observing runs, four of which spanned five nights 
in duration, with the final two points acquired through ESO Directors Discretionary Time.  All velocities and bisector span values (BVS) are listed 
in Table~\ref{tab:velocities}, note that the two large deviant BVS values are from our two lowest quality spectra which were interrupted by clouds, 
significantly lowering the S/N, meaning the construction of the CCF was much poorer than the rest of the high S/N spectra acquired.  Panel (a) in Fig.~\ref{fit} 
shows these velocities with the best fit Keplerian solution to the data.  Due to the scheduling of the runs the 
orbit is firmly constrained by the final two data points.  Before the final two data points were acquired, a resonant double planet 
solution could describe the dataset just as well.  Such system ambiguities could be common in radial-velocity datasets, 
particularly at low eccentricities (\citealp{anglada-escude08}).  However, in this case the system configurations were 
so contrasting that one data point was enough to decouple them, and with an eccentricity of around 0.63 and no indication of 
any trend in the third observational dataset, which is where a resonant system should most manifest a trend, we believe 
the quoted solution is accurate.  Panel (b) shows the residuals to this fit without correcting for the bisector variations.  Such 
variations have been shown to be a robust tracer of line asymmetries that can impact stellar radial-velocity measurements 
(\citealp{queloz01}; \citealp{henry02}).  We perform corrections for such variations following the analysis in \citet{migaszewski09} and 
the residuals for the data after BVS correction is shown in panel (c).  The radial-velocities (RV) are shown against the bisector velocity span (BVS) values in 
panel (d).  This hightlights the difference of the two offset velocities, particularly one data point which has a BVS value greater than 90m/s.  
The correlation coefficient ($\alpha$) used to correct the radial-velocities is shown in panel (e) by the solid line.  It is clear that the two velocities with large BVS values 
help to constrain the correlation across the parameter space and indicate the technique is robust, even for fairly large BVS offsets.  The $\alpha$ correction 
factor is found to be 0.697$\pm$0.064 and the corrected radial-velocities, those that are plotted in panel (a), are listed in column 5 of Table.~\ref{tab:velocities}.  
Clearly the data describes the fit much more accurately after 
BVS correction has been performed, compared with no corrections shown in panel (b).  Given the level of the residuals, with uncertainties 
which include in quadrature the stellar jitter of 1.50~m/s, being all around $\pm$1$\sigma$ from the fit, the quoted orbital solution is 
accurately constrained.  The orbital characteristics for a single companion solution is shown in column 2 of Table.~\ref{tab:mag_params}.

\begin{table}
\center
\caption{Radial-velocities for HD191760}
\label{tab:velocities}
\begin{tabular}{ccccc}
\hline
\multicolumn{1}{c}{BJD}& \multicolumn{1}{c}{RV} & \multicolumn{1}{c}{Uncertainty} & \multicolumn{1}{c}{BVS} & \multicolumn{1}{c}{RV-$\alpha$BVS} \\
\multicolumn{1}{c}{(-2454000)}& \multicolumn{1}{c}{(m/s)} & \multicolumn{1}{c}{(m/s)} & \multicolumn{1}{c}{(m/s)} & \multicolumn{1}{c}{(m/s)}  \\ \hline

250.90047&  -30662.99&   0.88&  9.49&      -235.23    \\
253.89490&  -30689.15&   0.70&  3.20&      -257.01    \\
365.50781&  -30575.97&   1.37&  9.45&      -148.19    \\
365.73781&  -30577.79&   0.62&  5.68&      -147.39    \\
367.49679&  -30540.86&   0.38&  3.66&      -109.04    \\
367.72155&  -30535.38&   0.73&  -0.09&     -100.95    \\
368.49981&  -30519.40&   0.53& -0.94&       -84.39    \\ 
368.71411&  -30498.29&   1.86&  30.70&      -85.31    \\ 
369.49468&  -30503.67&   0.77&  -1.71&      -68.12    \\
369.62706&  -30501.89&   0.50&  0.91&       -68.16   \\
369.68010&  -30497.67&   0.51&  2.50&       -65.07   \\
369.74838&  -30497.94&   0.63&  -0.93&      -62.93    \\
577.77658&  -30045.01&   0.82&  7.41&       384.19   \\
577.90446&  -30051.42&   0.40&  5.55&       379.07   \\
578.77151&  -30047.07&   0.54&  6.49&       382.77   \\
578.90713&  -30049.68&   0.48&  6.46&       380.18   \\
579.74685&  -30048.12&   0.56&  3.12&       384.07   \\
579.90348&  -30049.27&   0.41&  5.07&       381.57  \\
580.74769&  -30052.14&   0.60&  6.86&       377.44   \\
580.89930&  -30051.97&   0.51&  5.94&       378.25   \\
581.76210&  -30049.85&   0.39&  5.45&       380.71   \\
581.91331&  -30052.58&   0.36&  7.26&       376.73   \\ 
723.53335&  -30373.95&   3.17&  93.78&       -4.90   \\
724.55839&  -30440.36&   0.34& 7.46&        -11.19   \\
725.53823&  -30446.88&   0.44&  8.10&       -18.15   \\
726.53101&  -30452.77&   0.43&  7.76&       -23.81   \\
727.53134&  -30458.24&   1.09&  1.25&       -24.74   \\
812.52135&  -31646.47&   1.02&  12.07&    -1220.52   \\
813.53608&  -31679.70&   1.16&  10.27&    -1252.50   \\

\hline
\end{tabular}
\medskip
\end{table}

\section{Potential for Further Companion Objects}

In order to map out a strategy for future radial-velocity observations of this star, or indeed to decide if further study is warranted, we decided to look for 
additional hidden companions in the data.  This seems 
plausable given that the signal describes a fairly wide orbiting companion near the planetary mass boundary, therefore it is reasonable to assume there are other 
companions in the system with significantly lower mass that would induce a secondary orbital signature with a lower amplitude.  Therefore we performed a search 
for additional companion planets in the system even though the current dataset is rather limited.  

\begin{figure}
\vspace{5.0cm}
\hspace{-4.0cm}
\includegraphics{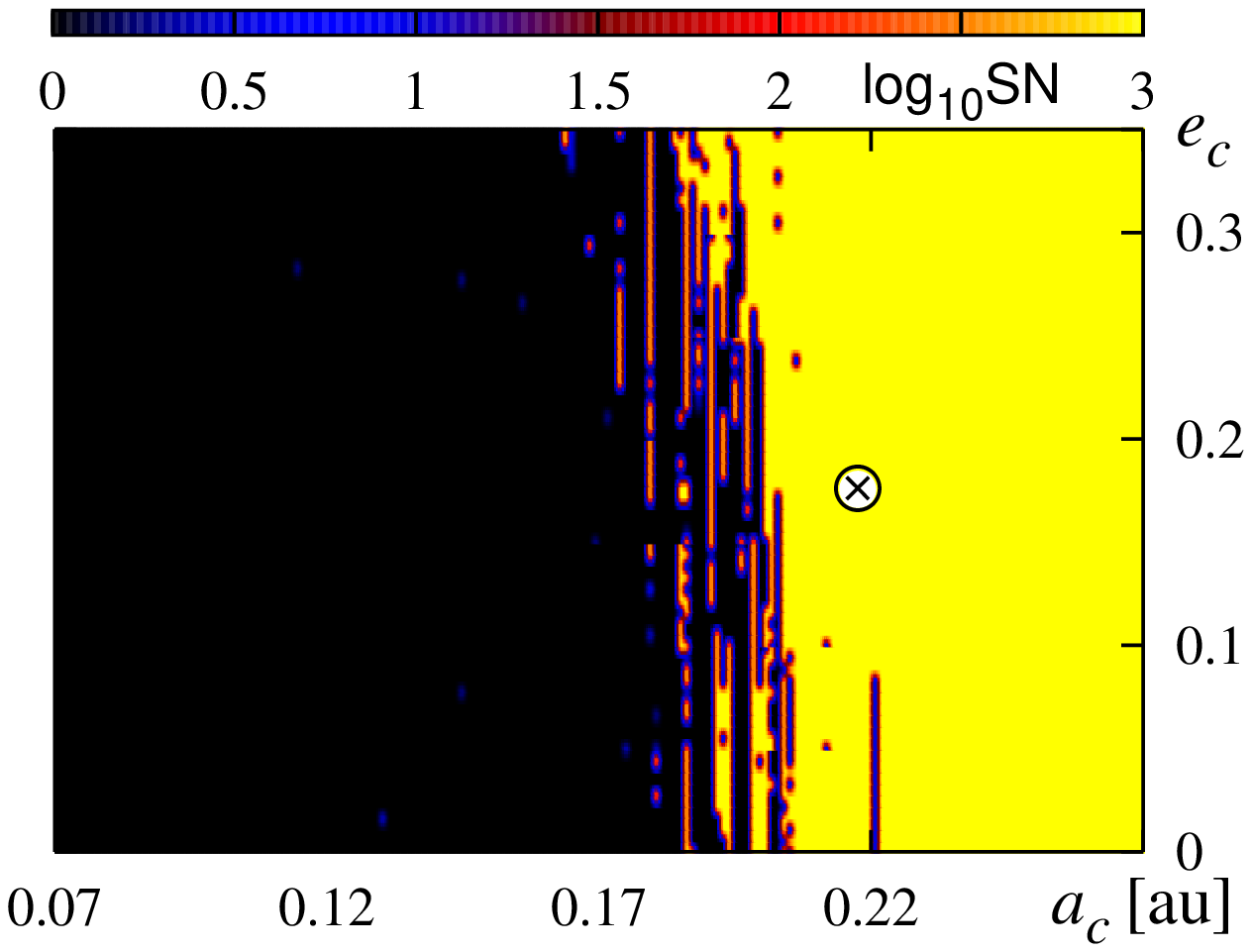}
\vspace{0.5cm}
\includegraphics{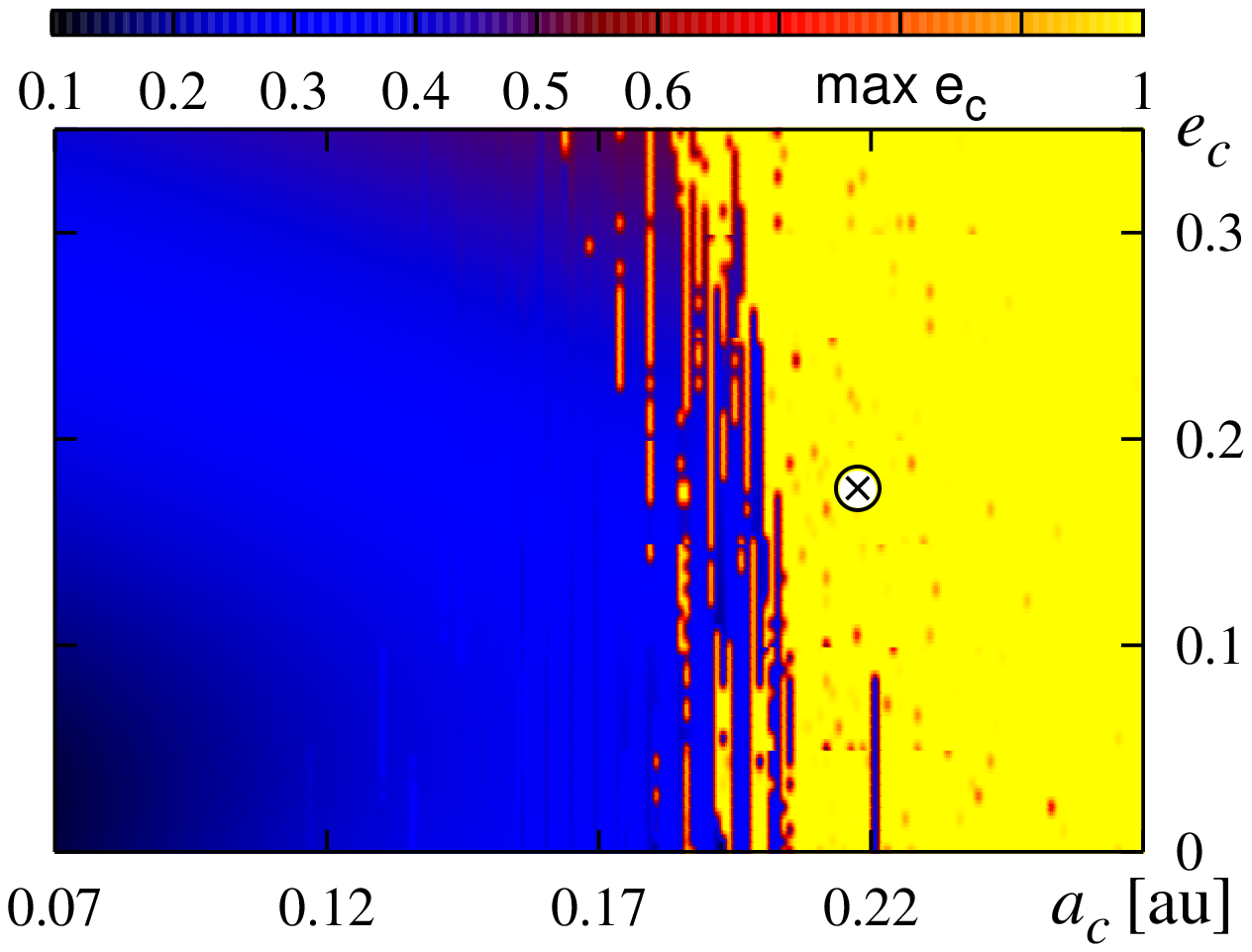}
\vspace{6.0cm}
\caption{Stability maps for a potential planet in semi-major axis--eccentricity space
in the HD191760 system.  A possible 2-planet solution of reasonable quality
and low eccentricity of the inner companion was found, and is shown by the
cross.  The top panel is for the Spectral Number indicator 
(\citealp{michtchenko01}), the lower panel is for the maximal
eccentricity attained by the putative planet during $10^5$~yrs of integration
time. The scale at the top of the panels represents the zones of stability,
with the darkest regions the most stable.  Both indicators around the
possible planetary object show no zones of stability anywhere. A sharp
border of stable motions is visible. Any planetary companion is unlikely to exist 
beyond 0.17--0.18~AU regardless of its initial eccentricity.}
\label{stability}
\end{figure}

First, we tested a hypothesis that the radial-velocity signal may be produced by
an additional planet with an intermediate period orbit. We performed the search for additional planets using the genetic
algorithm with MEGNO penalty (GAMP) (\citealp{gozdziewski03}; \citealp{gozdziewski06}). Essentially, this
numerical approach searches the phase space using a genetic algorithm to
find zones of stability where additional companions could reside. 
Unfortunately, the quite extensive GAMP search did not bring any reasonable
stable solutions beyond $\sim$0.2~AU limit.  Actually, we might expect that relatively far from the
star, only strongly resonant configurations could be found due to strong
gravitational influence of the brown dwarf.  We tested this by 
computing dynamical stability maps for reasonable fits found {\em without}
considering any stability checks. Such an example configuration returns a
secondary planetary object with a minimum mass of 0.10~M$_{\rm{J}}$, a low
eccentricity of $\sim$0.17 and a semimajor axis of $\sim$0.22~AU.  Such a configuration gives rise 
to a reduced rms of 1.84~m/s and could be
hidden in the current dataset.  However, as shown in Fig.~\ref{stability}
the entire region around the hypothetical planet is extremely chaotic and
therefore it is extremely unlikely that such a planet could reside at this location.  The two panels of
this figure represent stability maps in the semimajor axis-eccentricity
plane in terms of the so called Spectral Number indicator of formally chaotic motions (\citealp{michtchenko01},
top panel) and the maximum eccentricity indicator (the right-hand
panel), i.e. the maximum eccentricity attained during the integration period 
indicating the degree of geometric instability (large max e$_{c}$ means collisions).
Both maps were computed over $10^5$~years that corresponds to $\sim 50,000$ 
revolutions of the outer companion.  The figure shows that we must rule out, not only this
potential companion, but also all companions in the vicinity of this
parameter space. Actually, the strongly unstable zone ends at $\sim
0.17$--$0.18$~AU, and it is unlikely any other planet could survive the
disastrous, short-term perturbations of the brown dwarf companion beyond that limit.

\section{Discussion}

\subsection{Brown Dwarf Companion}

The constrained orbital parameters for HD191760$b$ place the minimum mass of the companion in the brown dwarf regime (M~sin~$i$=38.17$\pm$1.02M$_{\rm{J}}$).  Also the 
orbital semimajor axis is found to be 1.35~AU, well inside the brown dwarf desert region.  This so called brown dwarf desert was coined due to the paucity of 
such objects around solar-like stars with semimajor axes of $<$3-4~AU (\citealp{marcy00}).  They showed that fewer than 0.5\% of surveyed stars within 3~AU on Doppler 
programs at the time had brown dwarf companions, whereas 5\% of these stars harbour exoplanets.  Therefore, since our sample is small (100~stars) we may not 
statistically expect to detect any brown dwarf desert companions and 
this object may have an actual mass above the hydrogen burning limit ($\sim$80~M$_{\rm{J}}$) after inclination effects are considered.  For stars with randomly distributed inclinations, 68\% are found 
to fall within 47$^{\rm{o}}$ (\citealp{marcy96a}), giving rise to a true mass of M=52.19M$_{\rm{J}}$, comfortably below the hydrogen burning limit.  An inclination $i$ of, or below, 28.5$^{\rm{o}}$ is 
required in order for HD191760$b$ to be a true stellar companion and such a scenario has an $a$ $priori$ probability of only 12.1\%.  Also there is no binarity flag as of the latest update of the 
Hipparcos astrometry (\citealp{leeuwen08}).  In addition, a visual inspection of the HARPS spectra does not 
reveal any evidence for additional stellar lines from an unseen companion and there was no extension of the acquisition image at the telescope to indicate another stellar component, ruling out the 
possibility of extremely high inclinations.  However, such a scenario would agree 
with the results of \citet{stamatellos09} who suggest that brown dwarfs form by disk fragmentation in such systems and that short period radial-velocity companions are preferentially 
low mass stars.  This scenario would agree well with the high orbital eccentricity of the system and could also explain a recent similar discovery from \citet{wittenmyer09}.  Interferometry 
could be used to help rule out a lower mass stellar companion.

Another way to probe the inclination is to study the rotation velocity along with the activity index from the CaHK lines.  The rotation velocity 
of this star is found to be 2.33$\pm$0.05~km/s, using a similar method to that explained in \citet{santos02} (the uncertainty 
is the standard deviation of the 25 measured cross correlation full widths at half maximum), 
which for the determined radius we find gives an upper limit to the rotation period of 35~days.  The estimated rotation period from the 
log$R'_{\rm{HK}}$-rotation period correlation (\citealp{noyes}) is 25.2~days, which is inclination independent and should 
represent the equatorial rotation period, assuming that the orbital and rotation axes are aligned in 
the same plane.  Assuming the rotation period from the activity relation 
holds then we estimate the system has a sin~$i$=0.67, which gives an inclination of $\sim$42$^{\rm{o}}$.  Such an inclination would give rise to 
an actual mass of 57~M$_{\rm{J}}$, which is still comfortably below the hydrogen burning limit.  
However, the log$R'_{\rm{HK}}$ activity index was formulated for stars on the main sequence and may be gravity dependent so would not be fully 
applicable to this star.  Also the use of this index should be made only after a number of stellar 
cycles have been observed, which allows one a robust estimate of the mean log$R'_{\rm{HK}}$.  We 
have applied the value quoted in \citealp{jenkins08} (one measurement), not the mean, 
so we do not know the spread, which can vary at the $\pm$0.1~dex level (see \citealp{jenkins06c}), 
and can not apply this to the uncertainty on the relationship itself, which Noyes et al. quote as 0.08~dex for 
the convective turnover time.

\subsection{Benchmarkability?}

Since HD191760 is found to be evolving off the main sequence onto the sub-giant branch the age estimates for the system can be fairly well constrained.  \citet{holmberg07} estimate an age of 
5.6$^{+0.9}_{-1.9}$~Gyrs in good agreement with our estimated value of 4.1$^{+0.8}_{-2.8}$~Gyrs, the main differences arising due to the differing metallicities, and hence masses, found for this star.  
The mean of the 
two estimates gives rise to an age of 4.9$\pm$1.1~Gyrs, clearly a fairly old and evolved star.  Such an age means that this system could provide a benchmark, evolved system, to test brown dwarf 
evolutionary 
models.  Given there is a lack of well constrained age estimates for the older population of brown dwarf stars, if one can attain photometry or spectroscopy of the brown dwarf in this system then 
this could provide a well constrained metallicity and age to use in comparison with current evolutionary models of brown dwarfs (e.g. \citealp{baraffe03}; \citealp{burrows06}).  Again interferometry 
may allow such studies to be performed on this system.  Once the inclination is pinned down and the nature of the companion is confirmed as sub-stellar then we can also use the HARPS spectra 
to obtain a further age constraint from the strength of the lithium line (see \citealp{melo06}).  This would provide an additional, and independently determined, age estimate to further constrain the 
system.  In addition, further atomic abundances could be garnered for the star to aid in constraining the metallicity and abundance nature of brown dwarf model atmospheres.  Such an abundance study is 
already underway.

\section{Refined Orbits for the Planets Around HD48265, HD143361 and HD154672}

\begin{table}
\center
\caption{Radial-velocities for HD48265, HD143361 and HD154672}
\label{tab:mag_vels}
\begin{tabular}{ccc}
\hline
\multicolumn{1}{c}{BJD (-2454000)}& \multicolumn{1}{c}{RV (m/s)} & \multicolumn{1}{c}{Uncertainty (m/s)} \\ \hline
 &HD48265 &\\

365.824572	&23394.21	&3.47 \\
366.874164	&23399.40	&3.50 \\
367.852214	&23395.69	&3.45 \\
580.522907	&23449.18	&3.45 \\
581.553905	&23450.06	&3.45 \\
724.833698	&23413.42	&3.45 \\
725.817765	&23414.08	&3.44 \\
726.779101	&23411.87	&3.47 \\
\hline
 &HD143361 &\\

250.745318	&	-437.90&	2.39\\
253.769447	&	-436.44&	3.23\\
367.556628	&	-474.74&	2.22\\
368.537299	&	-474.20&	2.24\\
578.788567	&	-534.13&	2.18\\
581.807762	&	-531.75&	2.19\\
\hline
 &HD154672 &\\

248.713853&	-3796.59&	4.79 \\
250.731312&	-3794.42&	2.25 \\
367.595129&	-3653.29&	2.15 \\
368.618599&	-3651.00&	2.22 \\
578.811238&	-3797.81&	2.14 \\
581.870548&	-3808.85&	2.13 \\

\hline
\end{tabular}
\medskip
\end{table}

\begin{figure}
\vspace{2.25cm}
\hspace{-4.5cm}
\includegraphics{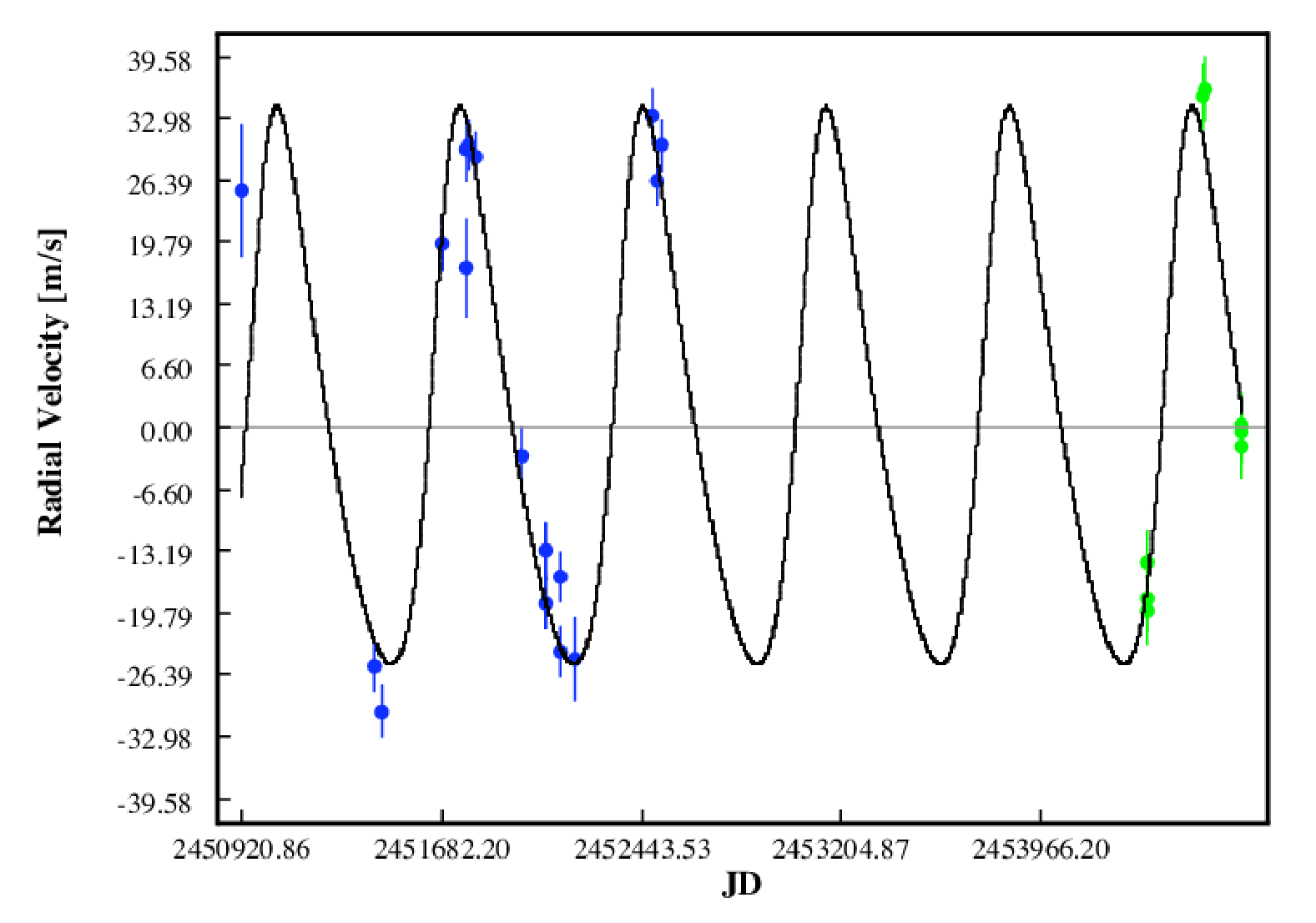}
\vspace{0cm}
\includegraphics{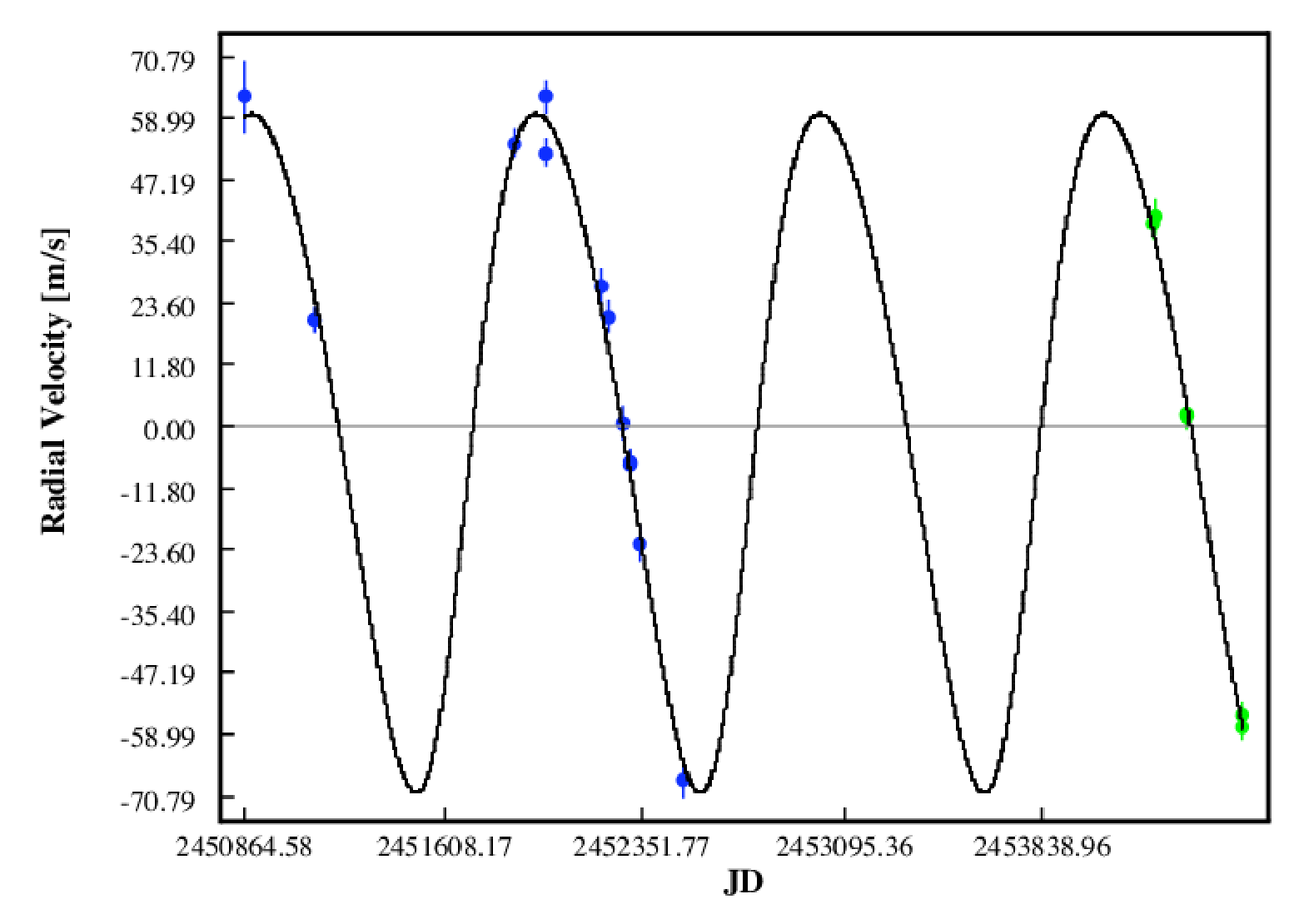}
\vspace{0cm}
\includegraphics{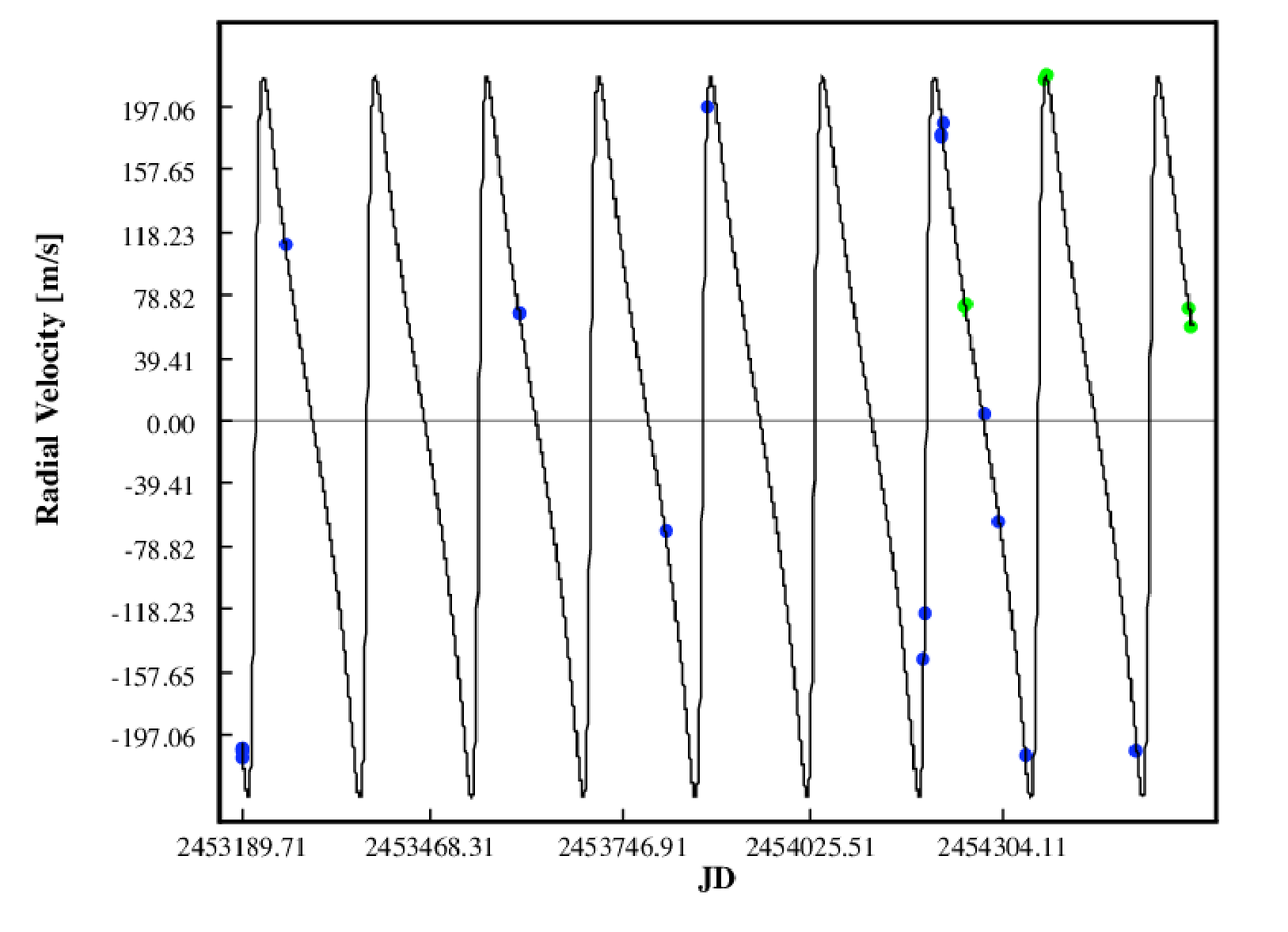}
\vspace{14.2cm}
\caption{The radial-velocity Keplerian fits to the stars HD48265 (top), HD143361 (middle) and HD154672 (bottom).  Plotted in blue are 
the literature values from the Magellan program and in green are the new data points from here.}
\label{mag_fits}
\end{figure}

Due to some overlap in our target sample with that of the Magellan Planet Search, we can update a few of the recent planet discoveries 
announced by this program with additional velocities we have found at HARPS.  Table.~\ref{tab:mag_vels} shows our absolute Dopper velocities for the three 
planet host stars HD48265, HD143361 and HD154672.  For all of these stars we have acquired observations in three observing runs with overall baselines 
around one year in duration.  Fig.~\ref{mag_fits} shows the best fits to all three stars using Systemic (\citealp{maschiari09}), with the Magellan velocities in black (blue) and our HARPS values 
in grey (green)\footnote{Colours in brackets are for the online version of the paper}.

We show the updated orbital solutions in columns 3,~4~and~5 of Table.~\ref{tab:mag_params} and in general they agree well with the published results in 
both \citet{minniti09} 
(HD48265 and HD143361) and \citet{lopez-morales08} (HD154672).  L{\'o}pez-Morales et al. used 16 velocity points in their orbital solution for HD154672 
over a number of orbital periods of the planet.  We find a very similar solution with a slightly reduced rms of 3.37~m/s.  
Note that we used the stellar masses quoted by this group for a direct comparison of the results found. 

For HD48265 we find fair agreement with the derived properties assuming a 1-planet Keplerian solution.  When adding our eight doppler velocities to the 17 acquired 
at Magellan, we find that the orbital period is now 700$\pm$8~days, albeit with a slightly increased rms of 6.78~m/s.  The $\chi$$^{2}_{\nu}$ also increases when 
compared to the Magellan data alone from a value of 1.37 up to 3.33.  Given the baseline between the two data sets this might argue for an additional companion in 
the system, however currently both a search in the periodogram of the residuals to the 1-planet fit and the addition of a linear trend do not reveal anything significant.  
Indeed, although this appears to be the best fit solution, there is no significant power anywhere in the periodogram when combining the two data sets.

Finally, for HD143361$b$ we find good agreement with the published results, again with a slightly reduced orbital period of 1057~days and eccentricity of 0.15.  This relates 
to an M~sin~$i$ of 3.12~M$_{\rm{J}}$, in good agreement with the 3~M$_{\rm{J}}$ found by Minniti et al. and with a lower rms of 3.37~m/s.  For a 1-planet solution none of these 
three systems exhibit significant trends in the residuals and hence present no evidence, as yet, for additional longer period planets in the system.  Note the 
uncertainties shown in Table.~\ref{tab:mag_vels} are derived from the updated relation of \citet{wright05} (private communication) and the low level of each is highlighted in 
Fig.~\ref{activity} where no Ca \sc ii \rm H line core emission is seen in any of these three stars (HD48265: top-middle, HD143361: lower-middle and 
HD154672: bottom).

\begin{table*}
\center
\caption{Orbital parameters for HD191760$b$, HD48265$b$, HD143361$b$ and HD154672$b$.}
\label{tab:mag_params}
\begin{tabular}{cccccc}
\hline
\multicolumn{1}{c}{Parameter} & \multicolumn{1}{c}{HD191760$b$} & \multicolumn{1}{c}{HD48265$b$} & 
\multicolumn{1}{c}{HD143361$b$} & \multicolumn{1}{c}{HD154672$b$}  \\ \hline

Orbital period $P$ (days)         & 505.65$\pm$0.42            & 700$\pm$8           & 1057$\pm$20      & 163.9$\pm$0.1\\
Velocity amplitude $K$ (m/s)      & 1047.83$\pm$38.71          & 28.3$\pm$9.0        & 65.1$\pm$26.3    & 226.52$\pm$9.22 \\
Eccentricity $e$                  & 0.63$\pm$0.01             &  0.18$\pm$0.13      & 0.15$\pm$0.17    & 0.61$\pm$0.02 \\
$\omega$ ($^{\rm{o}}$)             & 200.37$\pm$0.28           & 309$\pm$27          & 237$\pm$55        & 266$\pm$2 \\
$T$$_{\rm{0}}$ (JD-2,450,000)     & 4835.65$\pm$2.06          &  4486$\pm$50         &  3746$\pm$147     & 4520$\pm$1 \\
M~sin~$i$ (M$_{\rm{J}}$)             & 38.17$\pm$1.02               & 1.16$\pm$0.38       & 3.12$\pm$1.44    & 5.02$\pm$0.17\\
Semimajor axis $a$ (AU)           & 1.35                          &   1.51 	          & 2.00             & 0.60 \\
rms (m/s)                         & 2.00                         & 6.756      		   & 3.37             & 3.37 \\
$\chi$$^{2}_{\nu}$               & 1.04                         & 3.33      		   & 2.42             & 2.14 \\
N$_{\rm{Obs}}$                     & 29                          & 25$^{*}$                & 18               &  22  \\

\hline
\end{tabular}

\end{table*}

\section{Summary}

We announce the first four radial-velocity detections from the CHEPS on the ESO-HARPS instrument.  The highlight is the discovery of an eccentric brown dwarf in the desert orbiting the 
metal-rich G3IV-V star HD191760.  For the best fit solution, HD191760$b$ is found to have a mass of 38.17$\pm$1.02~M$_{\rm{J}}$, an orbital period of 
505.65$\pm$0.42~days and an eccentricity of 0.63$\pm$0.01, and this discovery adds to the small list of metal-rich stellar-brown dwarf binary systems.  Our dynamical simulations show that no additional 
planets could exist inside the orbit of the brown dwarf with semimajor axes larger than around 0.17~AU due to the destructive tidal forces of the brown dwarf.

Finally, we also confirm the exoplanet detections of HD48265$b$, HD143361$b$ and HD154672$b$ from the Magellan program by employing 1-planet Keplerian solutions to the data.  We show updated orbits 
by adding eight velocities for HD48265 and six velocities each for HD143361 and HD154672.  All orbital parameters are in agreement with the published 
results by \citet{minniti09} and \citet{lopez-morales08}.  Both HD143361$b$ and HD154672$b$ show lower rms by the addition of our data, however the best fit for HD48265$b$ shows an increased 
rms and $\chi$$^{2}_{\nu}$.  Since the star is shown to be chromospherically inactive this indicates that additional radial-velocities are required to probe for further companions in this system.

\chapter{\bf{Acknowledgements}}

We would like to acknowledge the help from both Ivo Saviane and Oliver Schuetz for observations carried out in the DDT program.  
J.S.J acknowledges partial support from Centro de Astrof\'\i sica FONDAP 15010003, along with partial support from GEMINI-CONICYT FUND and partial support 
from Comit\'e Mixto ESO-GOBIERNO DE CHILE.  K.G. and C.M. are supported by the Polish Ministry of Sciences and Education, Grant No. 1P03D-021-29.  We acknowledge the 
anonymous referee for their helpful comments.  We also 
acknowledge the Simbad and Vizier astronomical databases.  This research has made use of data obtained 
using the UK's AstroGrid Virtual Observatory Project, which is funded by the Science \& Technology Facilities Council and through the EU's Framework 6 
programme. 

\bibliographystyle{mn2e}
\bibliography{refs}

\label{lastpage}

\end{document}